\newcommand{\be}{\begin{equation}}
\newcommand{\ee}{\end{equation}}

\documentclass[twocolumn,floatfix]{revtex4}
\usepackage{graphicx}
\usepackage{dcolumn}
\usepackage{bm}
\usepackage{amsmath}

\begin{document}
\title{Nonlinear photonic lattices in anisotropic nonlocal self-focusing media}

\author{Anton~S.~Desyatnikov}
\author{Dragomir~N.~Neshev}
\author{Yuri~S.~Kivshar}
\affiliation{Nonlinear Physics Centre and Centre for Ultra-high
bandwidth Devices for Optical Systems, Research School of Physical
Sciences and Engineering, Australian National University,
Canberra, ACT 0200, Australia}

\author{Nina~Sagemerten}
\author{Denis~Tr\"{a}ger}
\author{Johannes~J\"{a}gers}
\author{Cornelia~Denz}
\affiliation{Institute of Applied Physics, Westf\"{a}lische
Wilhelms-Universit\"{a}t M\"{u}nster, D-48149 M\"{u}nster,
Germany}

\author{Yaroslav~V.~Kartashov}
\affiliation{ICFO-Institut de Ci\`{e}ncies Fot\`{o}niques and
Department of Signal Theory and Communications, Universitat
Politecnica de Catalunya, 08034 Barcelona, Spain}

\begin{abstract}
We analyze theoretically and generate experimentally
two-dimensional nonlinear periodic lattices in a photorefractive
medium. We demonstrate that the light-induced periodically
modulated nonlinear refractive index is highly anisotropic and
nonlocal, and it depends on the lattice orientation relative to
the crystal axis. We discuss stability of such induced photonic
structures and their guiding properties.
\end{abstract}


\maketitle

The study of nonlinear effects in periodic photonic structures
recently attracted strong interest because of many novel
possibilities to control light propagation, steering and trapping.
Periodic modulation of the refractive index modifies the linear
spectrum and wave diffraction and consequently strongly affects
the nonlinear propagation and localization of light~\cite{book}.

Photonic lattices can be optically induced by {\em linear}
diffraction-free light patterns created by interfering several
plane waves~\cite{Fleischer:2003-147:Nature}. However, the induced
change of the refractive index depends on the light intensity and,
in the nonlinear regime, it is accompanied by the self-action
effect~\cite{Chen:2002-2019:OL}. The nonlinear diffraction-free
light patterns in the form of stable self-trapped periodic waves
can propagate without change in their profile, becoming the
eigenmodes of the self-induced periodic potentials. This behavior
is generic, since {\em nonlinear periodic waves} can exist in many
types of nonlinear systems, and they provide a simple realization
of {\em nonlinear photonic crystals}. Such structures are
``flexible" because the lattice is modified and shaped by the
nonlinear medium; these flexible lattices extend the concept of
optically-induced gratings beyond the limits of weak material
nonlinearity. Moreover, the nonlinear lattices offer many novel
possibilities for the study of nonlinear effects in periodic
systems because they can interact with localized signal beams via
the cross-phase modulation and can form a composite bound
states~\cite{Desyatnikov:2003-153902:PRL,Neshev:2004-486:OL}.

Nonlinear photonic lattices created by two-dimensional arrays of
{\em in-phase} solitons have recently been demonstrated
experimentally in parametric
processes~\cite{Minardi:2000-1409:OL}, and in photorefractive
crystals with both coherent~\cite{Petter:2003-438:OL} and
partially incoherent~\cite{Chen:2002-2019:OL,Neshev:2004-486:OL,Martin:2004-123902:PRL}
light. For the case of two-dimensional arrays of in-phase solitons created by the amplitude modulation, every pixel of the lattice induces a waveguide which can be manipulated by an external steering beam~\cite{Petter:2003-438:OL, Martin:2004-123902:PRL, Petrovic:2003-R55601:PRE}. However, the spatial periodicity of
these lattices is limited by {\em attractive soliton interaction}
that may leads to their strong instability. In contrast, the
recently suggested two-dimensional lattices of {\em out-of-phase}
solitons are known to be robust in isotropic saturable
model~\cite{Kartashov:2003-015603:PRE}. The phase profile of such
self-trapped waves resembles chessboard with the lines of
$\pi$-phase jumps between neighboring white and black sites.

In this Letter, we study two-dimensional nonlinear lattices with a
chessboard phase structure in anisotropic nonlocal self-focusing
media and generate such lattices experimentally in a
photorefractive crystal. We demonstrate that the light-induced
periodically modulated nonlinear refractive index is highly
anisotropic and nonlocal, and it depends on the lattice
orientation relative to the crystal axis. We discuss stability of
such induced photonic structures and their guiding properties. An
obvious advantage of using this novel type of nonlinear periodic
lattices when compared with in-phase lattices or incoherent
soliton arrays~\cite{Chen:2002-2019:OL,Petter:2003-438:OL} is that
such lattices can be made robust even with smaller lattice
spacing. This creates an opportunity for experimental studies of
mutual coupling of periodic and localized waves recently discussed
for one-dimensional geometry, as well as the generation of
two-dimensional composite lattice
solitons~\cite{Desyatnikov:2003-153902:PRL}.

Spatially periodic nonlinear  modes appear naturally due to
self-focusing effect and modulational instability~\cite{book}.
When self-focusing compensates the diffraction of optical beams,
it may support both isolated spatial solitons and periodic soliton
trains, or stationary periodic nonlinear waves. The latter include
well studied {\em cnoidal waves}, described by the Jacobi elliptic
functions {\it cn} and {\it dn}, solutions to the nonlinear
Schr\"{o}dinger (NLS) equation,
\be \label{NLS} i \partial_z E+ \nabla^2 E + n(I) E= 0, \ee
where $I\equiv|E|^2$ and, for one-dimensional case,
$\nabla^2=\partial^2/\partial x^2=
\partial^2_x$. Similar stable periodic waves exist in different nonlinear models,
including quadratic, Kerr-type saturable nonlinearities etc. The family of
two-dimensional nonlinear periodic waves~\cite{Kartashov:2003-015603:PRE}
can be also extended to the case of the rectangular geometry with two different
transverse periods, because such anisotropic deformations of the square lattice
do not enhance its modulational instability. Stabilization of phase engineered
soliton arrays was reported recently for anisotropic model~\cite{Petrovic:2003-R55601:PRE}.

In this paper, we consider photorefractive crystal as an example
of anisotropic and nonlocal nonlinear media. In this case, the
nonlinear contribution to the refractive index in the model
(\ref{NLS}) is given by~\cite{Zozylya:1997-522:PRA}
\be \label{n}
n(I)=\Gamma \partial_x \varphi, \ee
where the electrostatic potential $\varphi$ of the
optically-induced space-charge field satisfies a separate
equation:
\be \label{pot} \nabla^2_{\perp} \varphi +\nabla_{\perp} \varphi
\nabla_{\perp} \ln(1+I)=\partial_x \ln(1+I). \ee
Here $\nabla_{\perp}$ is the two-dimensional gradient operator, $
\nabla^2_{\perp} =\partial^2_x + \partial^2_y$, and the intensity
$I$ is measured in units of the background illumination (dark)
intensity, necessary for the formation of spatial solitons in such
a medium. The physical variables $\tilde{x}$, $\tilde{y}$, and
$\tilde{z}$ correspond to their dimensionless counterparts as
$(\tilde{x},\tilde{y})=x_0(x,y)$ and $\tilde{z}=2\kappa x_0^2z$,
here $x_0$ is the transverse scale factor and $\kappa=2\pi
n_0/\lambda$ is the carrier wave vector with the linear refractive
index $n_0$. Parameter $\Gamma=x_0^2\kappa^2n_0^2r_{\rm
eff}\mathcal{E}$ is defined through the effective electro-optic
coefficient $r_{\rm eff}$ and externally applied bias
electrostatic field $\mathcal E$.

Stationary solutions to the system (\ref{NLS})-(\ref{pot}) are
sought in the standard form, $E(x,y,z)=U(x,y)\exp(ikz)$ where the
real envelope $U$ satisfies the equation
\be \label{stat} -kU+ \nabla^2_{\perp} U+\Gamma \partial_x\varphi U=0. \ee

We look for periodic solutions, $U(x,y) = U(x+2\pi,y+2\pi)$, and
solve Eqs.~(\ref{pot}), (\ref{stat}) using the relaxation
technique~\cite{Zozylya:1997-522:PRA} with initial ansatz in the
form of a linear periodic mode, $U_{\rm lin}(x,y) =A\sin x\,\sin
y$. We find that at least two distinct families bifurcate from the
linear wave $U_{\rm lin}(x,y)$, depending on its orientation: a
square pattern parallel to the {\it c}-axis, and a diamond pattern
oriented diagonally. Figures~1(a) and 1(b) show the field and
refractive index distributions in the low ($k=-1.9$, $A\simeq
0.9$) and relatively high ($k=-1.5$, $A\simeq 3.6$) saturation
regimes for both families. In a general case $\Gamma \neq 1$,
these two families occupy a band $k\in[-2,\Gamma-2]$ with the
amplitudes $A(k)$ and power densities $P(k)$ vanishing in the
linear limit $k\rightarrow -2$, see Fig.~1(c). Here the power
density is defined as the power of a unit cell,
$P=4\int\int_0^{\pi} U^2dxdy$. The main difference between two
solutions, clearly seen in Figs.~1(a) and 1(b), comes from the
refractive index: the regions with the effective focusing lenses
are well separated for the diamonds, and fuse to the effectively
one-dimensional stripes for square pattern, in the limit of strong
saturation. In Fig.~1(c), we plot maximal and minimal values
(Extrema) of the refractive index, Extr$(\partial_x\varphi)$.

\begin{figure}
\includegraphics[width=80mm]{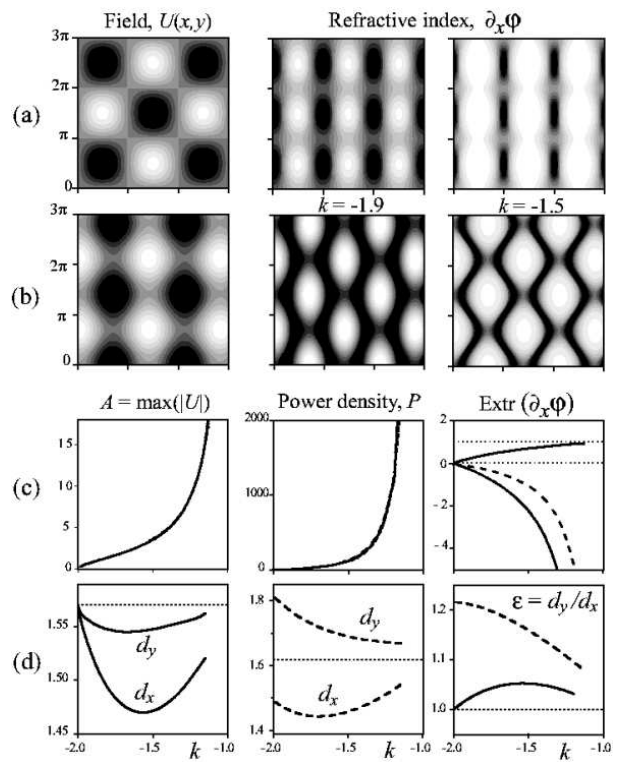}
\label{fig1}
\caption{(a) Square and (b) diamond self-trapped stationary periodic patterns
in the model (\ref{NLS})-(\ref{pot}) at $\Gamma=1$. Families of nonlinear
waves are summarized in (c) and (d); solid (dashed) lines correspond to
the square (diamond) patterns. Both lines for $A$, $P$, and $\max(\partial_x\varphi)$
coincide in (c). Horizontal dotted lines in (d) correspond to the linear limit $\Gamma\to 0$.}
\vspace{-5mm}
\end{figure}

In Fig.~1(d) we show the FWHM of a single intensity spot in two
orthogonal directions, $d_{x,y}$, characterizing the degree of
spatial localization of cnoidal
wave~\cite{Kartashov:2003-015603:PRE}. The ellipticity of every
lattice site, $\varepsilon = d_y/d_x\geq 1$, depends on the
propagation constant $k$, similar to the ellipticity of a single
photorefractive soliton~\cite{Zozylya:1997-522:PRA}, see
Fig.~1(d).

In order to test the lattice stability, we propagate numerically
two types of initially perturbed periodic solutions and observe
robust propagation for the distances exceeding the experimental
crystal length. Figure~2 demonstrates an example of a stable
propagation for the parameters close to our experimental
situation.

To demonstrate experimentally both existence and stability of
these nonlinear periodic lattices in anisotropic and nonlocal
media, we use a setup similar to that employed
earlier~\cite{Neshev:2004-486:OL}. Linearly polarized beam from a
frequency-doubled Nd:YAG laser at 532~nm is sent to a liquid
crystal programmable spatial light modulator in order to create a
periodic light pattern with a variable period and orientation. The
output of the modulator is then imaged by a high numerical
aperture telescope (demagnification $\sim 10$) on the front face
of a photorefractive SBN:60 crystal. The incident light is
linearly polarized parallel to the {\it c}-axis, thus experiences
strong photorefractive nonlinearity. The imposed pure phase
modulation transforms into an amplitude modulation of the beam at
the front face of the crystal, where noise is reduced by proper
spatial filtering. The crystal is externally biased and uniformly
illuminated with a white-light to control the dark irradiance. The
generated periodic pattern experiences robust linear propagation
in the crystal (at zero bias field) with a negligible change in
the periodicity. The linear output of the generated periodic wave
with periodicity of $31\mu$m at the back face of the crystal is
shown in Fig.~3~(left column) for two different orientations with
respect to the crystal axis.

\begin{figure}
\includegraphics[width=80mm]{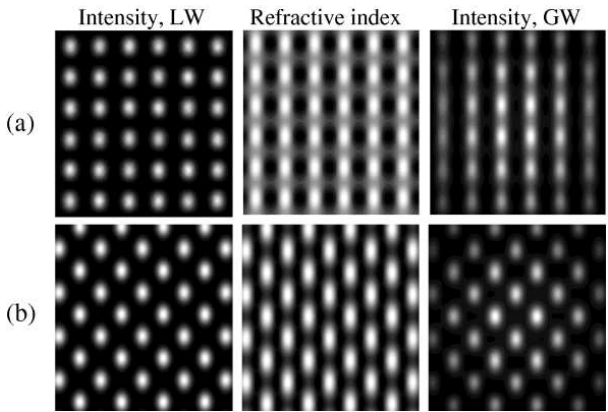}
\label{fig2} \vspace{-1mm} \caption{Numerical results for the propagation
of (a) square and (b) diamond self-trapped patterns for $\Gamma=11.8$ in
the low saturation regime $A\approx 1$, $k=-0.5$. Intensities of the lattice
and probe wave it guides are shown after propagation $\tilde{z}\approx 23$~mm.
On the input, the lattice is perturbed with $20$\% of random noise, and the
probe is a broad Gaussian beam.}
\vspace{-2mm}
\end{figure}

\begin{figure}
\includegraphics[width=80mm]{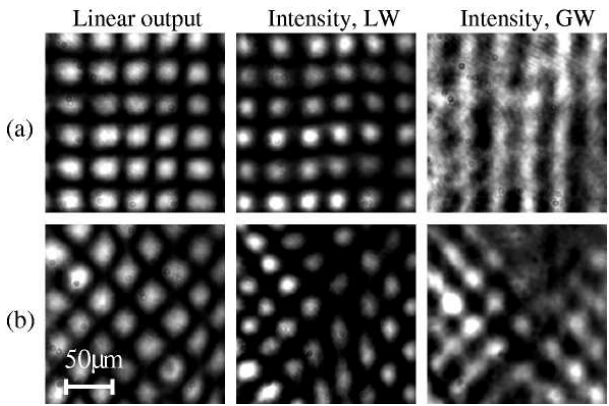}
\label{fig3} \vspace{-1mm} \caption{Experimental results for the (a) square
and (b) diamond lattices. Left to right: output intensities of the lattice after
linear and nonlinear propagation, and output of a guided plane wave, respectively.}
\vspace{-2mm}
\end{figure}

Applying an external DC electric field (${\mathcal E}=1000$ V/cm)
across the crystal creates appropriate conditions for the
formation of spatial solitons and, correspondingly, influences the
propagation of the periodic waves. Since with increasing of the
nonlinearity (i.e., biasing voltage) the output of the lattice
does not change significantly [Fig.~3~(middle column)], the only
way to test that the lattice propagates nonlinearly is to probe
the periodic potential it induces. To do so, we send a broad plane
wave trough the crystal and observe its modulation at the output.
In practice, this is realized by switching-off the voltage on the
modulator, thus removing the modulation imposed on the light
pattern and generating a broad plane-wave at the input of the
crystal. This approach assures that the plane wave is propagating
exactly along the induced waveguides. Due to a slow response of
the photorefractive crystal, we can quickly monitor the output of
the plane wave without modifying the induced refractive index
change. Output intensity distributions for two orientations of the
lattice pattern are shown in Fig.~3~(right column). The plane-wave
gets split into several channels, and we observe guiding of the
probe beam at the maxima of the refractive index. The experimental
pictures shown in Fig.~3 are in a good agreement with the
corresponding numerical simulations of the nonlinear anisotropic
nonlocal model presented in Fig.~2, and demonstrate a qualitative
difference of the guided patterns for two different orientations
of the lattice.

In conclusion, we have studied theoretically and generated
experimentally two-dimensional nonlinear photonic lattices in
anisotropic photorefractive medium. We have found two distinct
classes of self-trapped robust spatially-periodic waves with
out-of-phase neighboring sites, the square pattern oriented
parallel to the crystal axes, and the diamond pattern oriented
diagonally in the transverse plane. We have demonstrated that the
highly anisotropic refractive index distribution induced by the
lattice differs significantly from its isotropic counterpart and
depends strongly on the lattice orientation.

This work was supported by the Australian Research Council and the
Alexander von Humboldt Foundation.

\end{document}